\documentstyle[prl,eqsecnum,multicol,epsfig,aps]{revtex}   

\begin{document}

\draft

\title{Intrinsic Tunneling in Cuprates and Manganites}

\author{S. Heim, T. Nachtrab, M. M\"o{\ss}le, and R. Kleiner}
\address{Physikalisches Institut-Experimentalphysik II,
Universit\"at T\"ubingen, Auf der Morgenstelle 14,\\ D-72076 T\"ubingen, Germany}

\author{R. Koch, S. Rother, O. Waldmann, and P. M\"uller}
\address{Physikalisches Institut III,
Universit\"at Erlangen-N\"urnberg, Erwin-Rommel-Stra{\ss}e 1,\\ D-91058 Erlangen, Germany}

\author{T. Kimura, and Y. Tokura}
\address{Joint Research Center for Atom Technology (JRCAT),\\ Tsukuba 305-0046, Japan}

\date{July 23, 2001}

\maketitle

\begin{center}
\begin{minipage}{16cm}
\begin{abstract}
The most anisotropic high temperature superconductors like Bi$_{2}$Sr$_{2}$CaCu$_{2}$O$_8$, as well as the recently
discovered layered manganite La$_{1.4}$Sr$_{1.6}$Mn$_2$O$_7$ are layered metallic systems where the interlayer current
transport occurs via sequential tunneling of charge carriers. As a consequence, in Bi$_{2}$Sr$_{2}$CaCu$_{2}$O$_8$
adjacent CuO$_2$ double layers form an intrinsic Josephson tunnel junction while in in La$_{1.4}$Sr$_{1.6}$Mn$_2$O$_7$
tunneling of spin polarized charge carriers between adjacent MnO$_2$ layers leads to an intrinsic spin valve effect. We
present and discuss interlayer transport experiments for both systems. To perform the experiments small sized mesa
structures were patterned on top of single crystals of the above materials defining stacks of a small number of
intrinsic Josephson junctions and intrinsic spin valves, respectively.
\end{abstract}
\end{minipage}
\end{center}


\begin{multicols}{2}



\section{Introduction}

\begin{figure}
\epsfig{file=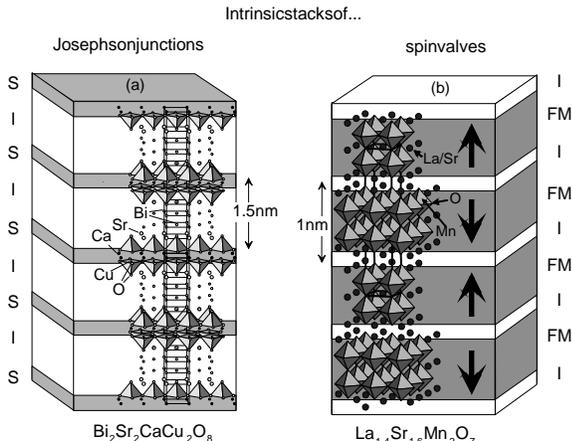, width=7.5cm}
\caption{High temperature superconductor Bi$_2$Sr$_2$CaCu$_2$O$_8$ intrinsically forming a stack of Josephson junctions (a) and layered manganite La$_{1.4}$Sr$_{1.6}$Mn$_2$O$_7$ intrinsically forming a stack of spin valves (b). S denote superconducting, FM ferromagnetic and I insulating layers. Arrows in (b) denote direction of magnetization at low temperatures.}\label{fig01}
\end{figure}

It is well established that the most anisotropic high temperature superconductors form superconducting multilayers
where the superconducting unit (single layers, bi- or trilayers of CuO$_2$) are separated by insulating barrier
layers. Examples are Bi$_{2}$Sr$_{2}$CaCu$_{2}$O$_8$ (BSCCO) or Tl$_{2}$Ba$_{2}$Ca$_{2}$Cu$_{3}$O$_{10}$. Interlayer
transport occurs via sequential tunneling of charge carriers, and consequently suitably patterned structures can be
considered as a stack of intrinsic Josephson tunnel junctions (cf. Fig.~\ref{fig01}a). The intrinsic Josephson effect \cite{Kle92,Yur00} has been intensively investigated within the last decade for a number of reasons. For example, the gap voltage limiting the high frequency properties of intrinsic Josephson junctions is in the range of 30 mV allowing ac Josephson currents at frequencies of several THz \cite{Rot01}. Thus, such junctions are highly promising for THz applications like oscillators or mixers. Second, a number of fundamental phenomena can be investigated like nonequilibrium effects due to charge imbalance in the superconducting layers \cite{Mac99,Pre98} or collective effects like Josephson plasma oscillations involving the stack as a whole \cite{Sak94,Bul94}, the collective dynamics of Josephson fluxons \cite{Kle94,Kle01}, or the formation of charge solitons \cite{Lat99}. A third issue is to use
intrinsic Josephson junctions for tunneling spectroscopy. Their advantage over artificial tunnel junctions is to probe
not only the surface layer but all layers within the stack. The prize, however, is that always a number of layers are
measured in series. Thus the information provided is essentially the average over several layers. Using this method
phonons in Bi- and Tl- based cuprates have been detected \cite{Hel97,Sch98}. Recently, the method has been used to
investigate the quasiparticle density of states in both the superconducting and normal regime \cite{Suz99,Kra01}.

A system very similar to the high temperature superconductors are the layered manganites like
(La,Sr)$_{3}$Mn$_{2}$O$_{7}$ \cite{Mor96,Kim96}. Here, MnO$_2$ bilayers are separated by thin layers consisting of La
and Sr ions. In La$_{1.4}$Sr$_{1.6}$Mn$_2$O$_7$ (LSMO) the MnO$_2$ bilayers undergo a metal-insulator transition near
90\,K while the intervening layers remain insulating. Within each MnO$_2$ sheet magnetic moments are ferromagnetically
ordered. As revealed by neutron diffraction measurements \cite{Per98} adjacent sheets become antiferromagnetically
arranged below 70\,K with magnetization vector perpendicular to the layers. Interlayer transport of the spin-polarized
charge carriers occurs via tunneling processes \cite{Kim96,Per98}, leading to a low temperature tunneling
magnetoresistance (TMR) in addition to the effect of Colossal Magnetoresistance (CMR) observed near the
metal-insulator transition (cf. Fig.~\ref{fig01}b). Thus, at low temperatures, adjacent MnO$_2$ bilayers can be considered as a natural TMR element or spin valve switching from large to small tunneling resistance upon application of an external
magnetic field. Consequently, a suitably patterned single crystal acts as an intrinsic stack of spin valves
\cite{Nac01}.

What is a suitable size of an intrinsic Josephson junction stack or an intrinsic spin valve stack to be investigated?
In terms of intrinsic Josephson junctions in BSCCO there are two limitations. First, the Josephson length
${\lambda}_{J}$ determining, e.g., the size of a Josephson fluxon is in the order of 0.2 - 0.5\,$\mu$m
\cite{Kle92,Yur00,Kle94,Kle01}. If the formation of such fluxons is to be avoided all lateral dimensions of the stack
should not exceed ${\lambda}_{J}$. In case of a "long" Josephson stack still the smaller side of the stack should be
below ${\lambda}_{J}$. The second requirement is that the  stack should not contain too many junctions. For many
fundamental investigations stacks of well below some tens of junctions are desirable in order to produce interpretable
results. Also, for stacks containing a large number of junctions ohmic heating becomes severe. The thickness of the stack
thus should be below some 10\,nm. In terms of the layered manganites hall probe measurements \cite{Fuk99} and
magneto-optical investigations \cite{Wel99} have revealed that LSMO single crystals exhibit a large number of magnetic
domains typically some 10\,$\mu$m in size. In order to observe clear spin valve effects, structures smaller than the
size of such domains are desirable.  For the same reasons as for intrinsic Josephson junctions also here the stacks
should consist of only a few unit cells.

\section{Samples and Measurements}
BSCCO single crystals were grown from a stoichiometric mixture of the oxides and carbonates \cite{Ger91}. Epoxy was used
to glue approximately $1 \times 1\ \times 0.1\,\mu$m$^2$ large crystals to a sapphire substrate. To obtain a
sufficiently small contact resistance the crystals were cleaved immediately before mounting them into the vacuum
chamber and the crystal surface was covered with silver. The contact resistance was typically $10^{-5}\,\Omega$cm$^2$.
Subsequently, rectangular mesa structures with minimal lateral dimensions of 0.5\,$\mu$m were patterned using
electron beam lithography and argon-ion milling. For electrical insulation of the lead contacting the top of the mesa a
250\,nm thick SiO layer was evaporated. The top contact was provided by a 300-400\,nm thick gold or silver layer.
Currents were extracted from the base crystal using large pads contacting its top surface. The leads and contact pads
were patterned by photolithography and argon-ion milling. The LSMO single crystals of roughly 1\,mm$^3$ in size were
grown using a floating zone technique \cite{Mor96,Kim98}. Mesa structures were patterned using the same techniques as
for the BSCCO single crystals. The only difference was that the crystal surface was initially polished mechanically
instead of being cleaved. The mesa size ranged from $5 \times 5$ to $10 \times 10\,\mu$m$^2$. Their thickness was about 20\,nm corresponding to a stack of 20 spin valves. For both systems transport measurements were performed in a two-terminal configuration. Low pass filters were used to reduce external noise and the bias current was provided by a battery powered current source.

\section{Results}
\subsection{Intrinsic Josephson junctions in Bi$_{2}$Sr$_{2}$CaCu$_{2}$O$_8$}

In this section we will discuss some results obtained for intrinsic Josephson junction stacks. The main focus will be
on quasiparticle tunneling in both the superconducting and the normal state.
\begin{figure}
\epsfig{file=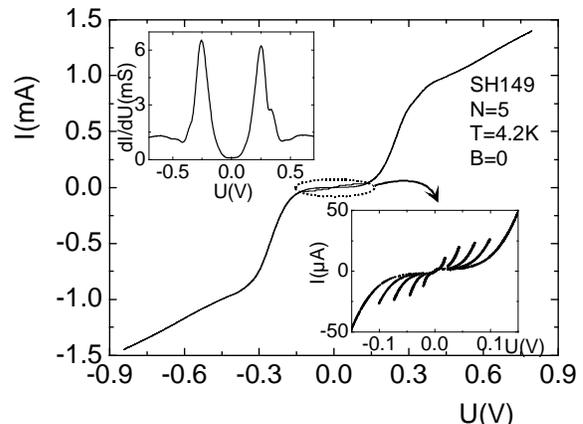, width=7.5cm}
\caption{Current voltage characteristic at 4.2\,K of $1 \times 1\,\mu$m$^2$ BSSCO mesa structure SH149 containing 5 junctions. Lower inset shows same characteristic on expanded scale, upper inset shows derivative $dI/dU$.}\label{fig02}
\end{figure}

\begin{figure}
\epsfig{file=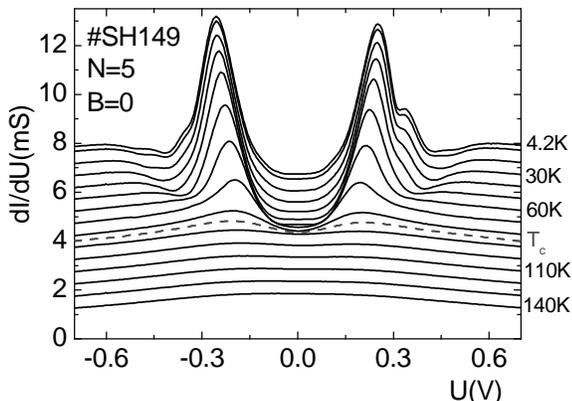, width=7.5cm}
\caption{Conductance $dI/dU$ of BSCCO mesa SH149 at temperatures between 4.2\,K and 140\,K. From 10\,K to 140\,K temperature is increased in steps of 10\,K. in addition curve at $T_c$ is shown. Curves are vertically offset in steps (140\,K - T)/20.}\label{fig03}
\end{figure}

Fig.~\ref{fig02} shows a typical current voltage ($I$-$U$) characteristic of $1 \times 1\,\mu$m$^2$ large mesa SH149 patterned on a slightly overdoped BSCCO single crystal with $T_{c} = 86\,$K, as determined from the onset of interplane
superconductivity. The mesa consisted of N = 5 junctions. Below the critical currents of these junctions the $I$-$U$
characteristic exhibits 5 branches in the resistive state differing by the number of resistive junctions (lower inset
of Fig.~\ref{fig02}). On the large current scale all junctions are resistive. The upper inset in Fig.~\ref{fig02} shows the conductance $dI/dU$. A clear gap structure is visible with a total gap value $2N\Delta = 0.25$\,V (the sum of the gap voltages of all junctions) corresponding to $\Delta = 25$\,meV. Below the gap the conductance is U-shaped. Also note the dip and hump feature above $2\Delta$ typical for BSCCO tunneling spectra \cite{Tim99}. In Fig.~\ref{fig03} we show conductance curves of the same mesa at various temperatures. While the amplitude of the gap peak strongly decreases and the dip and hump features disappear when approaching $T_{c}$, the voltage position of the gap changes only weakly and almost continuously transits into the pseudogap regime. 

\begin{figure}
\epsfig{file=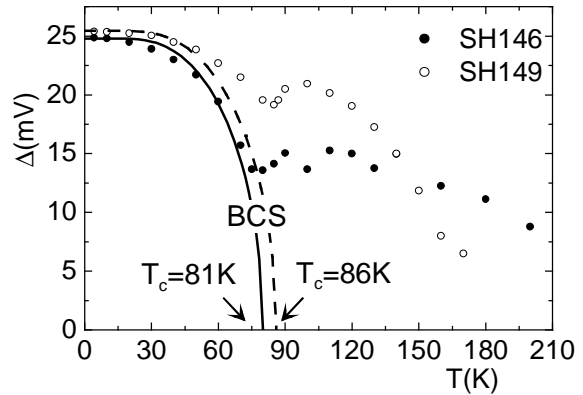, width=7.5cm}
\caption{Peak voltage (divided by $2N$) of conductance curves of conductance curves vs. $T$ for the two slightly overdoped BSCCO mesas SH149 ($T_c=86$\,K) and SH146 ($T_c=81$\,K). Lines correspond to BCS temperature dependence of the gap.}\label{fig04}
\end{figure}

\begin{figure}
\epsfig{file=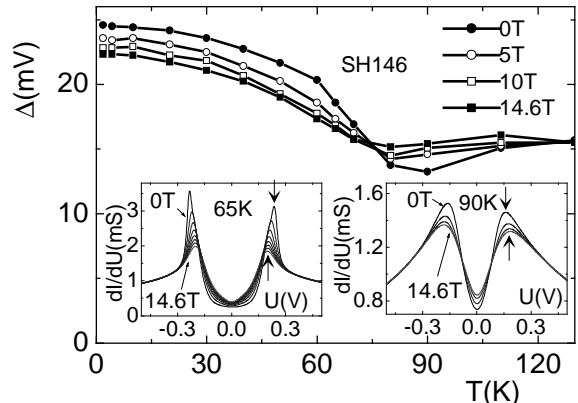, width=7.5cm}
\caption{Peak voltage (divided by $2N$) of conductance curves of BSCCO mesa SH146 vs. temperature for several values of magnetic fields perpendicular to the layers. Insets show conductance curves for $T=65$\,K (left) and $T=90$\,K (right). Vertical arrows in insets denote peak positions for $B=0$ and $B=14.6$\,T. In left inset field is increased in steps of 2\,T, in right inset in steps of 5\,T.}\label{fig05}
\end{figure}

This can be seen more clearly in Fig.~\ref{fig04} where we plot $\Delta$ vs. $T$ for mesa SH149 as well as for 6 junction $0.8 \times 0.8\,\mu$m$^2$ mesa SH146. Also this mesa was slightly overdoped with a $T_{c}$ of 81\,K. For both mesas $\Delta(T)$ exhibits a minimum near $T_{c}$. While in the superconducting state $\Delta$ is almost identical for both mesas they differ strongly in both magnitude and temperature dependence of $\Delta$ above $T_{c}$. When applying a magnetic field perpendicular to the layers we found that $\Delta$ decreases in the superconducting state while it is almost field independent or even increases slightly with field in the pseudogap regime. 
Fig.~\ref{fig05} illustrates this via curves $\Delta(T)$ of mesa SH146 for various values of magnetic field up to 14.6\,T. Particularly, the minimum in $\Delta$ near $T_{c}$ is levelled out. The insets of this figure show a series of conductance curves for fields between 0 and 14.6\,T in the superconducting state at $T = 65$\,K (left inset) and in the normal state at $T = 90$\,K (right inset). Vertical arrows mark the field dependent shift of the conductance maxima.

An important question is to what extent the two phenomena gap and pseudogap are related. To our opinion, the intrinsic
tunneling data suggest they are not for several reasons. In contrast to the gap in the superconducting state the
pseudogap is strongly material dependent. The dip in $\Delta(T)$ near $T_{c}$ even gives the feeling that two phenomena
compete with each other. In terms of magnetic field dependence gap and pseudogap clearly behave differently. We also
investigated slightly underdoped samples where we found that $\Delta$ almost goes to zero at $T_{c}$. For these samples
gap and pseudogap were observable simultaneously in the superconducting state. Although the above features - similar
observations have also been made by other groups investigating intrinsic tunneling  in BSCCO \cite{Suz99,Kra01} - are
certainly not a proof of the independence or even competition of two unrelated phenomena, they at least raise doubts of
a common mechanism leading to the superconducting gap and to the pseudogap.

\subsection{Intrinsic spin valves in La$_{1.4}$Sr$_{1.6}$Mn$_2$O$_7$}

Next we turn to the properties of mesa structures patterned on LSMO single crystals. In contrast to (transport)
measurements on bulk single crystals the mesa technique provides the possibility to probe a small region well below the
size of a magnetic domain. We will show that such mesas indeed exhibit the properties expected for a stack of intrinsic
spin valves.

\begin{figure}
\epsfig{file=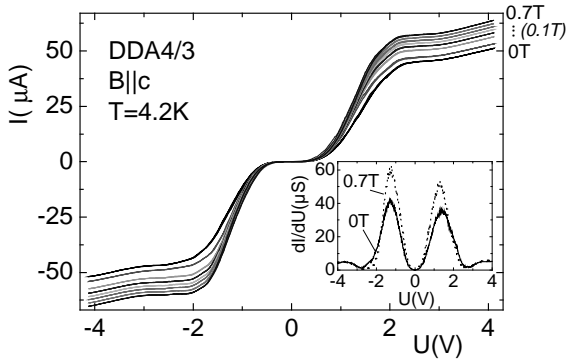, width=7.5cm}
\caption{Current voltage characteristics of $5 \times 5\,\mu$m$^2$ large mesa DDA4/4 patterned on a LSMO single crystal in magnetic fields between 0 and 0.7\,T. The field, oriented perpendicular to the layers, is increased in steps of 0.1\,T. Inset: derivative $dI/dU$ for $B=0$ (solid line) and for $B=0.7$\,T (dashed line)}\label{fig06}
\end{figure}
Fig.~6 shows field dependent 4.\,K $I$-$U$ characteristics of the $5 \times 5\,\mu$m$^2$ large mesa DDA4/4 having a
thickness of about 20\,nm. The resistance is clearly lowered with increasing magnetic field. The inset shows the
conductance $dI/dU$ for $B = 0$ and $B = 0.7$\,T. There is a conductance maximum at 1.4\,V for $B = 0$, corresponding to roughly 75\,mV per spin valve (with a 50\,\% error margin, since we do not know the number of layers in the mesa precisely)
slightly shifting to 1.3\,V for $B = 0.7$\,T. The overall conductance curves look strikingly similar to the low temperature
tunneling spectra of BSCCO mesas (cf. Fig.~\ref{fig02}) although the similarities might be accidental. A possible reason for the dip structure in the LSMO mesa might be the excitation of spin waves \cite{Kat00}. We also note that similar
conductance curves as in Fig.~\ref{fig06} are obtained for temperatures up to the metal-insulator transition, with a slight
($\sim 10-20\,\%$) decrease in the conductance maximum.

What can be expected for the low temperature magnetoresistance of a mesa probing a single magnetic domain? In zero
field, the magnetization vectors of adjacent MnO$_2$ bilayers are aligned antiparallel (cf. Fig.~\ref{fig01}). In magnetic fields parallel to the layers the magnetization will tilt continuously towards parallel orientation leading to a continuous
decrease of tunneling resistance. In perpendicular fields there should be a spin flop from antiparallel to parallel for
a field where the Zeeman energy ($\propto HM$) overcomes the interlayer coupling energy ($\propto M^2$). Consequently,
there should be a discontinuous jump in magnetoresistance for perpendicular fields.

\begin{figure}
\epsfig{file=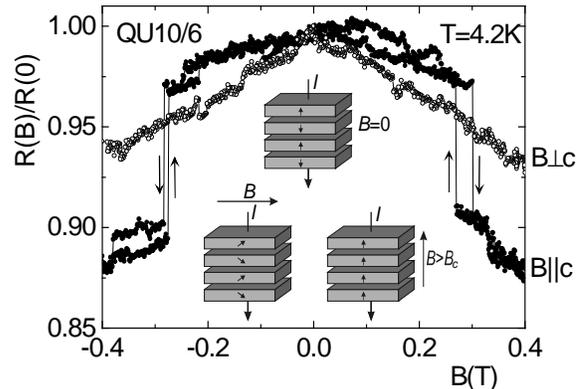, width=7.5cm}
\caption{Magnetoresistance of LMSO mesa QU10/6 for field orientation parallel (solid symbols) and perpendicular (open symbols) to the layers. Bias current is set in the subgap regime of the current voltage characteristic. Inset illustrates magnetization vectors for zero field, parallel field and perpendicular field above the threashold field where the jump in $R(B)$ occurs.}\label{fig07}
\end{figure}
Fig.~\ref{fig07} shows data for the $10 \times 10\,\mu$m$^2$ large mesa QU10/6 for both field orientations. In parallel fields $R(B)$ indeed decreases continuously while in perpendicular fields one jump near 0.3\,T is observed with a small
hysteresis of about 30\,mT for increasing/decreasing fields. We observed similar data for in total 10 mesas. In
perpendicular fields the jump in magnetoresistance was observed in fields between 50\,mT and 0.4\,T, without apparent
correlation to mesa size or thickness. The observed range of switching fields corresponds nicely to magnetooptical
measurements on LSMO single crystals where first order spin flop transitions have been observed between 0.11 and 0.48\,T. From the fact that only one single switching event is visible in $R$ for perpendicular fields we conclude that all
spin valves within the stack switch collectively (i.e. the magnetization vectors of all MnO$_2$ bilayers flip
simultaneously). Note, for comparison, that intrinsic Josephson junctions in BSCCO can be switched to the resistive
state one by one. For a more detailed discussion of the magnetoresistance observed for our mesa structures, see
\cite{Nac01}.

We finally return to the $I$-$U$ characteristics of the LSMO mesa structures. While the $I$-$U$ characteristics shown in Fig.~\ref{fig06} were single-valued and the voltage changed continuously with bias current multiple hysteresis were observed when the bias current was increased above some threshold. Here, very similar to the $I$-$U$ characteristics of intrinsic Josephson junction stacks, hysteretic branches appeared. We observed such behavior for all 10 mesa structures investigated so far. 
\begin{figure}
\epsfig{file=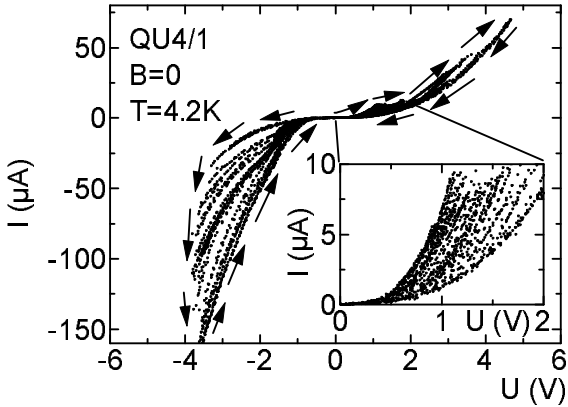, width=7.5cm}
\caption{Current voltage characteristic of $10 \times 10\,\mu$m$^2$ LSMO mesa QU4/1 exhibiting multiple hysteresis. Arrows indicate direction of current sweeps.}\label{fig08}
\end{figure}
As an example, Fig.~\ref{fig08} shows an $I$-$U$ characteristic of the $10 \times 10\,\mu$m$^2$ mesa Qu4/1 showing the multiple hysteresis in a very pronounced way. Increasing the current $I$ from zero one observes for $I<10\,\mu$A a continuous branch which stays stable as long as the current does not increase about $+10\,\mu$A or, at negative bias, the voltage does not exceed $-3.5$\,V. Increasing the current above $+10\,\mu$A leads to a jump towards larger voltage, and a new continuous branch is observed. For negative bias a jump towards smaller voltage occurs when the voltage across the
stack is below $-3.5$\,V. Sweeping the current between jumps at positive and negative bias many times yielded the $I$-$U$
characteristic shown in Fig.~\ref{fig08}. Arrows in this figure denote one possible sweep yielding parts of the innermost and outermost branches. We observed this behavior in fields up to 7\,T and also for temperatures up to at least 90\,K where
the branches started to become instable. Since at tesla fields the magnetization of the crystal has already fully
saturated we consider it unlikely that the observed hysteresis are due to switching of individual spin valves. We may
observe some layer by layer charge ordering effects known to exist in layered manganites. However, a clear explanation
of the multiple hysteresis still needs to be found.

\section{Conclusions}

The above data may have shown that there are a number of similarities between the cuprate
Bi$_{2}$Sr$_{2}$CaCu$_{2}$O$_8$ (BSCCO) and the layered manganite La$_{1.4}$Sr$_{1.6}$Mn$_2$O$_7$ (LSMO). In both
systems interplane transport occurs via sequential tunneling of charge carriers. Suitably prepared mesa structures on
BSCCO and LSMO single crystals act as stacks of intrinsic Josephson tunnel junctions and spin valves, respectively.
Both systems allow the investigation of a number of electrodynamic effects like Josephson fluxon dynamics or the
collective switching of spin valves.

From a microscopic point of view there is a striking similarity between the tunneling spectra of both systems at low
temperatures. Further investigations will show to what extent a comparison between the two materials will lead to an
improved understanding of both the high temperature superconductors and the manganites.

\section{Acknowledgments}

Financial support by the Deutsche Forschungsgemeinschaft and the Bayrische Forschungsstiftung is gratefully
acknowledged.


%
%

\end{multicols}


%
%




\end{document}